# *Emerging trend in the east-west Dipole Pattern in Indian Summer Monsoon Rainfall and the associated impact on Regional Dynamics*

*Akshara Satheesh[1] and Rajib Chattopadhyay*[1]*

*Indian Institute of Tropical Meteorology, Pune-411008*

**Corresponding Author email: rajib@tropmet.res.in*

# Abstract

The Indian Summer monsoon is a dominant climatological feature that is responsible for rainfall across most regions of India, and it exhibits significant variabilities across multiple spatiotemporal scales. Traditionally, during the monsoon season, more rainfall is received along the Western Ghats, the Northern Gangetic plains, the central belt, and northeast India. However, recently, there has been a shift in this canonical monsoon rainfall pattern on the monthly to seasonal scale. In this study, we quantify an east-west asymmetric trend in monthly to seasonal rainfall due to the increased rainfall over the northwestern part of the country. An Empirical Orthogonal Function (EOF) analysis has been performed to understand the spatial and temporal variation of the monsoon. EOF mode 3 shows such a distinct east-west dipole pattern, highlighting the existence of a modal feature representing the recent trend in the rainfall distribution. The physical nature of this mode is also established. The regression pattern of the rainfall anomalies to the Webster-Yang Index (Webster and Yang, 1992) exhibits a similar east-west pattern that further confirms the physical existence of this east-west rainfall modal dipole pattern. Since rainfall across the northwest is directly linked to the Arabian Sea and rainfall over the eastern region to the Bay of Bengal, the characteristics of these two regions are studied separately. Over the Arabian Sea, there is a significant negative trend in the Sea Level Pressure (SLP) anomalies and an increase in the specific humidity, causing greater moisture convergence. In contrast, over the Bay of Bengal, the SLP shows an increasing trend. The SST warming over the Arabian Sea is higher than that of the Bay of Bengal. Further, while investigating the zonal wind(u) at 850hPa, it shows an increasing trend along the northern branch that is more directed towards the northwestern part of the country. These factors together create dynamically favorable conditions for enhanced convection and thus receive more rainfall across the northwest compared to the northeast India.

# 1. Introduction

The Indian summer monsoon exhibits pronounced variability across both spatial and temporal scales. The Monsoon rainfall availability varies from year to year; even when the total seasonal rainfall remains the same, the spatial distribution may vary. These variabilities can be attributed to the complexity of the Indian monsoon system, which is influenced by a combination of global-scale atmospheric-oceanic interactions such as ENSO (Webster et al. 1998; Goswami 1998; Krishnamurthy & Goswami, 2000), the Indian Ocean Dipole (IOD) (Ashok et al. 2001; Ashok et al. 2004, and other factors such as Atlantic-Nino Atlantic Niño/AMO (Nagaraju et al., 2018). Many local factors also have a significant influence on the Indian Summer Monsoon.

Climatologically, during the summer monsoon season, high rainfall occurs along the west coast of the peninsula (associated with orography parallel to the coast) and over the northeastern regions. In addition, there is a broad zone around 20±N, stretching northwestward from the head of the Bay of Bengal, which receives significant rainfall, known as the monsoon zone (Gadgil, 2003;Sikka and Gadgil 1980). The dominant Empirical Orthogonal Function (EOF) pattern of the Indian monsoon also represents its classic active-break pattern of the Indian monsoon.

However, in recent years, including the current year 2025, a shift in the canonical monsoon rainfall distribution has been easily noticeable. **Fig.1a** and **b** show the rainfall anomalies for the June to September (JJAS) monsoon season and for July and August, respectively. Similarly, plots for rainfall anomalies for many other recent years (2010, 2023,2024) are shown in the other panels of **Fig.1**. These plots show that more rainfall has been received in northwest India than in northeast India during recent years. Several studies and meteorological have supported this observation. For example, the recent 2024 Indian Monsoon Report (Mohapatra et al., 2025) by the India Meteorological Department (IMD), indicated increased rainfall over the western regions, particularly over Saurashtra and Kutch in Gujarat, which marked an increasing trend in what was traditionally considered a relatively arid zone. In contrast, there was a declining rainfall trend in eastern and northeastern India. Together, this amplifies the east-west precipitation gradient across the subcontinent.

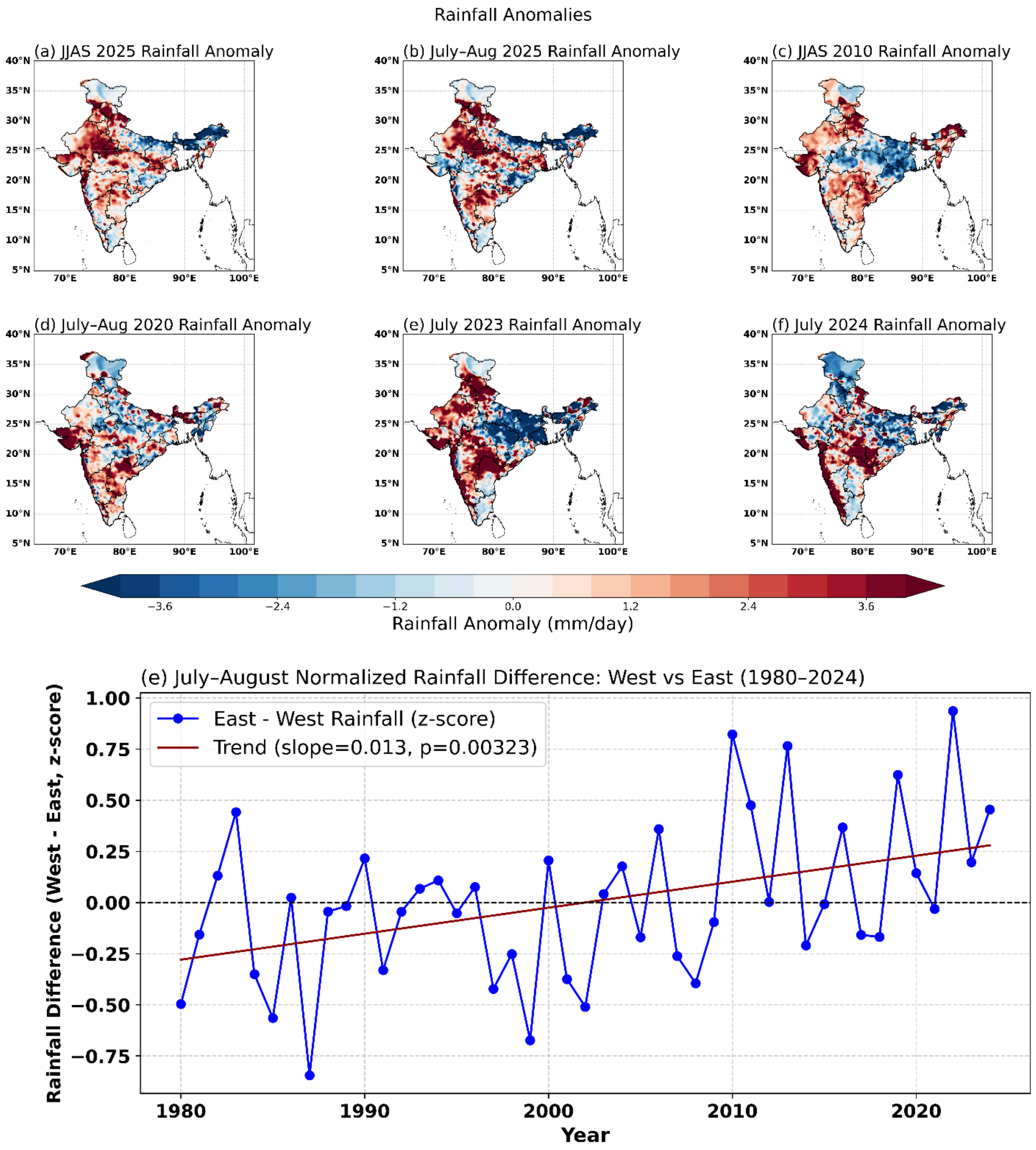

**Fig.1** Rainfall anomalies obtained from IMD datasets for (a) JJAS season 2025 (b) July-August 2025 (c) JJAS season 2010 (d) July-August 2020 (e) July 2023 (f) July 2024 (g) annual variability of west-east gradient of standardized rainfall anomaly (The red line shows the linear trend).

Recent studies have suggested a significant expansion of the Indian monsoon system to the north and west. A study from the Council on Energy, Environment and Water (https://www.ceew.in/) on the tehsil level reported that statistically significant JJAS rainfall increases were observed in the traditionally drier tehsils of Rajasthan, Gujarat, central Maharashtra, and parts of Tamil Nadu, whereas decreased rainfall is witnessed over the Indo-Gangetic plains that contribute to more than half of India's agricultural production, northeastern India, and the Indian Himalayan region (Prabhu and Chitale, 2024). These observations show spatial inhomogeneity in the monsoon rainfall distribution.

The spatial inhomogeneity of the rainfall distribution is a well-known problem. We refer the reader to **Fig.1** again. **Fig.1** (b - d) illustrates the spatial inhomogeneity of rainfall anomaly patterns of 2010 (June-September JJAS average) – received normal seasonal rainfall, 2020 (July–August) – received near-normal rainfall, 2023 (July) received above normal rainfall, and 2024 (July) – received above normal rainfall. These years were randomly selected to understand monsoon rainfall distribution on a monthly to seasonal scale. This shows that in recent years, positive anomalies have emerged over northwest India, including Gujarat, Rajasthan, and parts of Maharashtra, while negative anomalies appear mainly across the Gangetic plains, including Bihar, Uttar Pradesh, and parts of West Bengal, indicating a clear shift in monsoon rainfall distribution. This broadly points to east-west inhomogeneity and is quantified further in **Fig. 1(e),** which shows the yearly variation of an east-west rainfall gradient. It was calculated by subtracting the area-averaged standardized rainfall anomaly over the eastern region (20–27°N, 80–89°E) from the western region (20–27°N, 71–80°E). This clearly shows a steepening west-to-east rainfall gradient, especially after 2010, which shows a clear positive gradient. This further validates the presence of a significant spatial shift in the monsoon rainfall patterns.

Monsoon rainfall supplies up to 80 % of the annual precipitation in India and affects the livelihoods of more than a billion people. Therefore, the timing and intensity of monsoon rainfall play important roles in determining the country's economy ( Gadgil, S. and Gadgil, S. 2006; Turner & Annamalai, 2012). This emerging pattern challenges the conventional understanding of monsoon rainfall distribution and also impacts the water resource management, agriculture, and disaster preparedness across India.  Therefore, understanding the mechanisms driving rainfall redistribution is critical for refining seasonal forecasting models and devising adaptive strategies for climate resilience.

This study mainly focuses on explaining this emerging east-west shift in Indian Summer Monsoon Rainfall on a monthly to seasonal scale. It is concentrated mainly in the core monsoon months of July-August. The study also explored the large-scale atmospheric dynamical features and surface oceanic features associated with this shift. The regional redistribution of monsoon rainfall influences agriculture, water resources, and socio-economic conditions across the country. Therefore, it is crucial to understand the driving mechanisms of this shift to improve seasonal prediction and develop region-specific adaptation strategies under a changing climate.

# 2. Data and Methodology

## 2.1 Data

This study used multiple datasets to examine the monsoon variability and its associated components. It includes (a) the daily gridded precipitation data from the Indian Meteorological Department (IMD), with horizontal resolution of 0.25˚×0.25˚ (Pai et al., 2014) (b) Monthly precipitation data from the Global Precipitation Climatology Project dataset with horizontal resolution 2.5˚×2.5˚ (GPCP; Adler et al., 2018 ) (c) monthly reanalysis data of zonal and meridional wind and Sea Level Pressure (SLP) is obtained from NCEP-NCAR reanalysis data set having horizontal resolution 2.5˚×2.5˚ (Kalnay et al., 1996). (d) Monthly sea surface temperature (SST) data were sourced from the Hadley Centre Sea Ice and Sea Surface Temperature dataset with a horizontal resolution 1.0˚x 1.0˚ (HadISST; (Rayner et al., 2003)), NOAA Extended Reconstructed SST V5 data with resolution of 2.0˚ x 2.0˚ (Huang et al., 2017) and COBE-SST 2 and Sea Ice data of resolution 1.0˚x 1.0˚ (Hirahara et al., 2014). The analysis was conducted on a monthly scale from July to August for the period 1980 - 2024

## 2.2 Methodology

To understand the dominant patterns of rainfall variability on a monthly scale, an Empirical Orthogonal Function (EOF) analysis was performed on the IMD rainfall anomalies. In this study, rainfall and other variables were calculated according to the climatology of the period 1980 - 2024 and used throughout the study. Simple linear regression analysis was conducted to understand the influence of circulation patterns on monthly rainfall anomalies. Regression of rainfall anomalies performed with respect to the Webster-Yang Index. Webster-Yang index (WYI) is defined as the vertical zonal (u) wind shear between 850hPa and 200hPa ($U_{850}$-$U_{200}$) between 0˚-20˚N latitude and 40˚-110˚E longitude (Webster and Yang, 1992).

An east-west rainfall gradient index was constructed to understand the shift in the rainfall pattern. Here, the area-averaged standardized rainfall anomaly amplitude over two different regions was taken: an eastern region(Ae) between 20–27°N latitude and 80–89°E longitude, and a western region($A_w$) between 20–27°N latitude and 71–80°E longitude. This index is defined as:

$$\mathbf{A_{we}=A_w - A_e}.$$

Additionally, Probability Density Function (PDF) analysis of rainfall anomalies and associated factors was performed to quantify and detect their shift in the recent period (2003 - 2024) compared to an earlier time period (1980 - 2002). To better understand the shift in regional dynamical drivers, the Arabian Sea region (5˚-20˚ N, 57˚-72˚E) and Bay of Bengal (5˚-20˚ N, 80-95˚ E) were studied separately. For examining SST warming, the Arabian Sea domain was defined as 6–24°N latitude and 54–72°E longitude, and the Bay of Bengal domain as 5–23°N latitude and 80–96°E longitude, in order to exclude land areas and maximize ocean grid coverage. This integrated approach provides a comprehensive assessment of rainfall variability, its dynamic and thermodynamic drivers, and evolving characteristics of monsoon rainfall over India.

# 3. Results and Discussions

## 3.1 Spatial Patterns of Monsoon Rainfall and their relations with the large-scale dynamical features

To identify the leading modes of the Indian rainfall anomaly for July-August on a monthly time scale, dominant modes of rainfall based on EOF analysis were calculated using IMD and GPCP precipitation data sets (supplementary). The two datasets exhibit similar spatial and temporal distributions. **Fig.2** shows the first three dominant EOF modes of Indian Monsoon Rainfall for

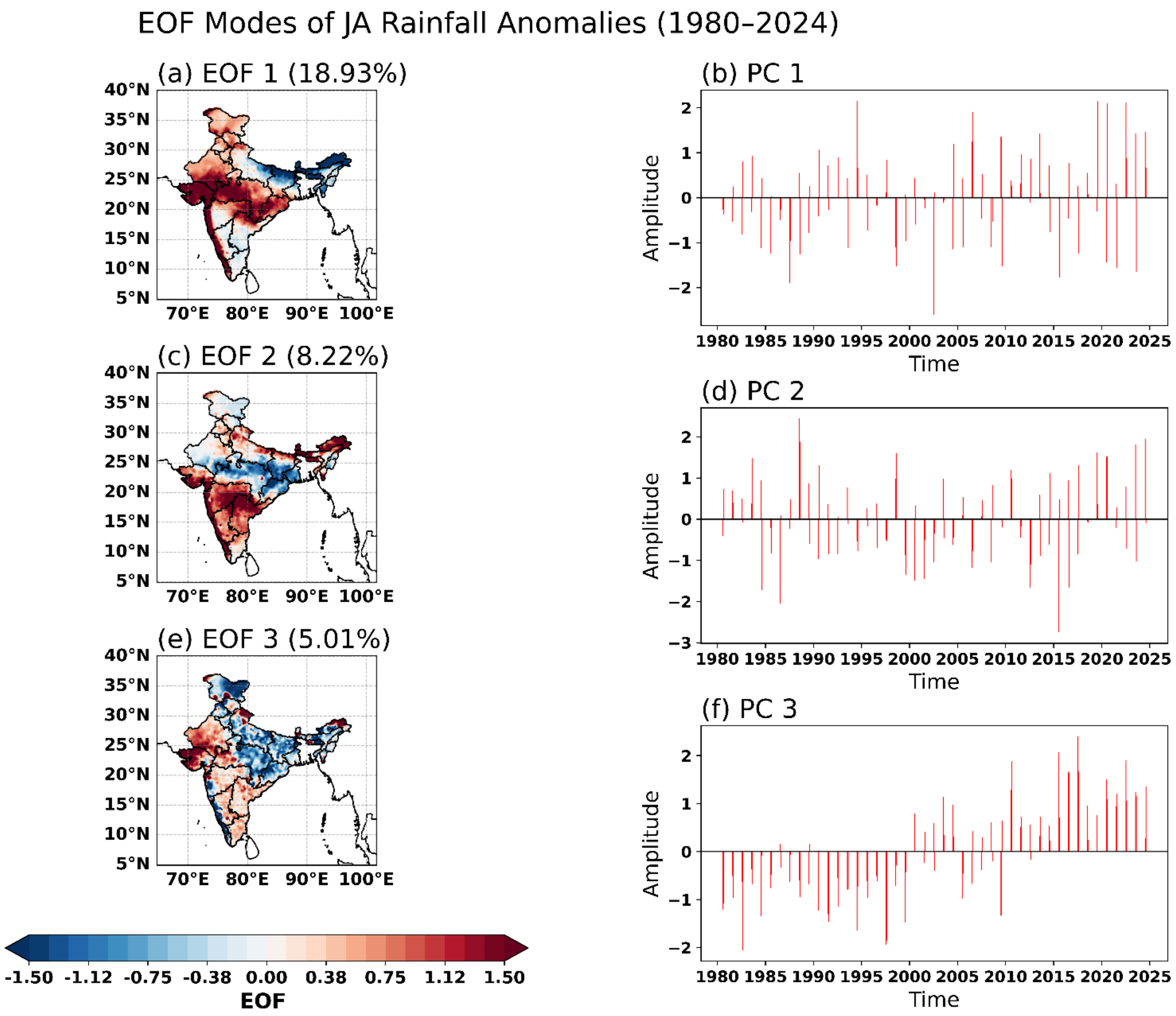

**Fig. 2** First 3 dominant modes of rainfall EOF pattern (left) and their corresponding PCs (right) of monthly rainfall anomalies for July-August derived from the IMD data. The percentage of variance is given on the top-right corner of EOF plot.

July-August. Here, the first mode explaining a variance of 18.86% (**Fig.2a**) shows a spatially uniform pattern over central India, the Western Ghats, southeastern India, and the Himalayan foothills. This represents the classic seasonal mean and active-break pattern of the Indian monsoon, capturing the spatial variability associated with monsoon phases (Ramamurthy 1969; Krishnamurthy & Shukla, 2000). In fact, the variations in all-India summer monsoon rainfall are highly correlated with the variation in rainfall over this monsoon zone represented by the EOF1 pattern (Sikka and Gadgil, 1980). Therefore, changes in the associated systems of the monsoon also affect this dominant mode. Therefore, the reported decrease in the magnitude of the tropical jet stream (Abish et al., 2013; Sreekala et al., 2014), which is an important component of the Indian Summer Monsoon (Koteswaram, 1958),can inversely affect the monsoon rainfall and result in reduced rainfall over the central belt of India, especially over the northeast.

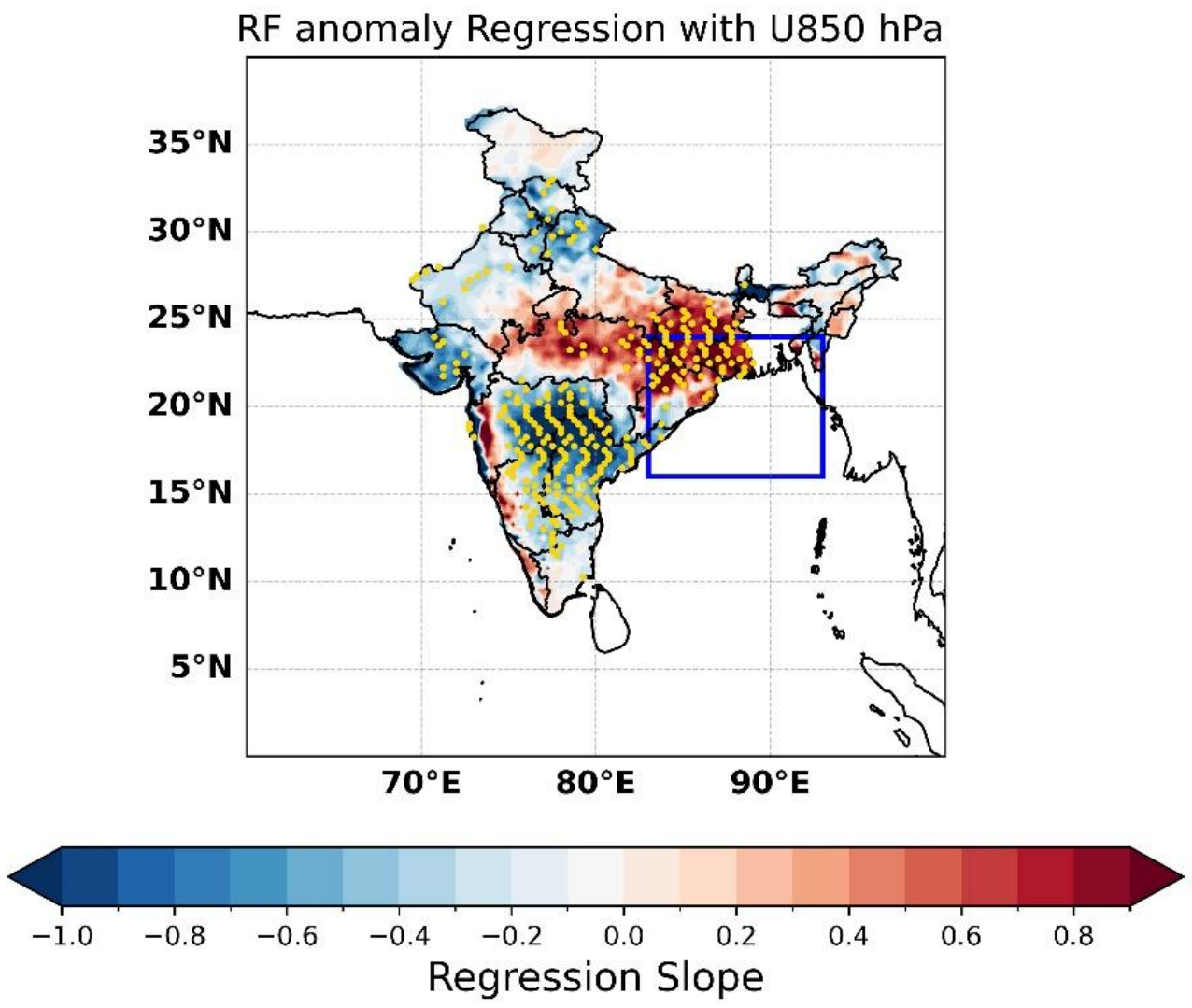

**Fig.3** Regression map of Rainfall anomaly with respect to a reference index of zonal wind(u) at 850hPa. The reference index is created from u850hPa-wind is averaged over the region 15-25˚N latitude and 83 to 93˚E longitude for July-August season. The marked region(yellow) represents the statistically significant region and the blue box represents the area where u-wind is considered.

The second mode of EOF analysis represents the north-south rainfall variability along the central belt of the country, lying along the regions of the monsoon trough, and explains a variance of 8.13% (**Fig.2 (c)**). Our analysis suggests that this mode is mainly associated with the circulation patterns over the Bay of Bengal. **Fig. 3** shows the regression pattern of the monsoon rainfall with zonal wind at 850hPa across the head of the Bay of Bengal between 15-25˚N latitude and 83 to 93˚E longitude, which clearly resembles mode 2 of the EOF analysis. Furthermore, PC2 and zonal wind over the Bay of Bengal have a correlation of 0.35 which is statistically significant based on a Student's t-test indicated an association between mode 2 and the Bay of Bengal circulation.

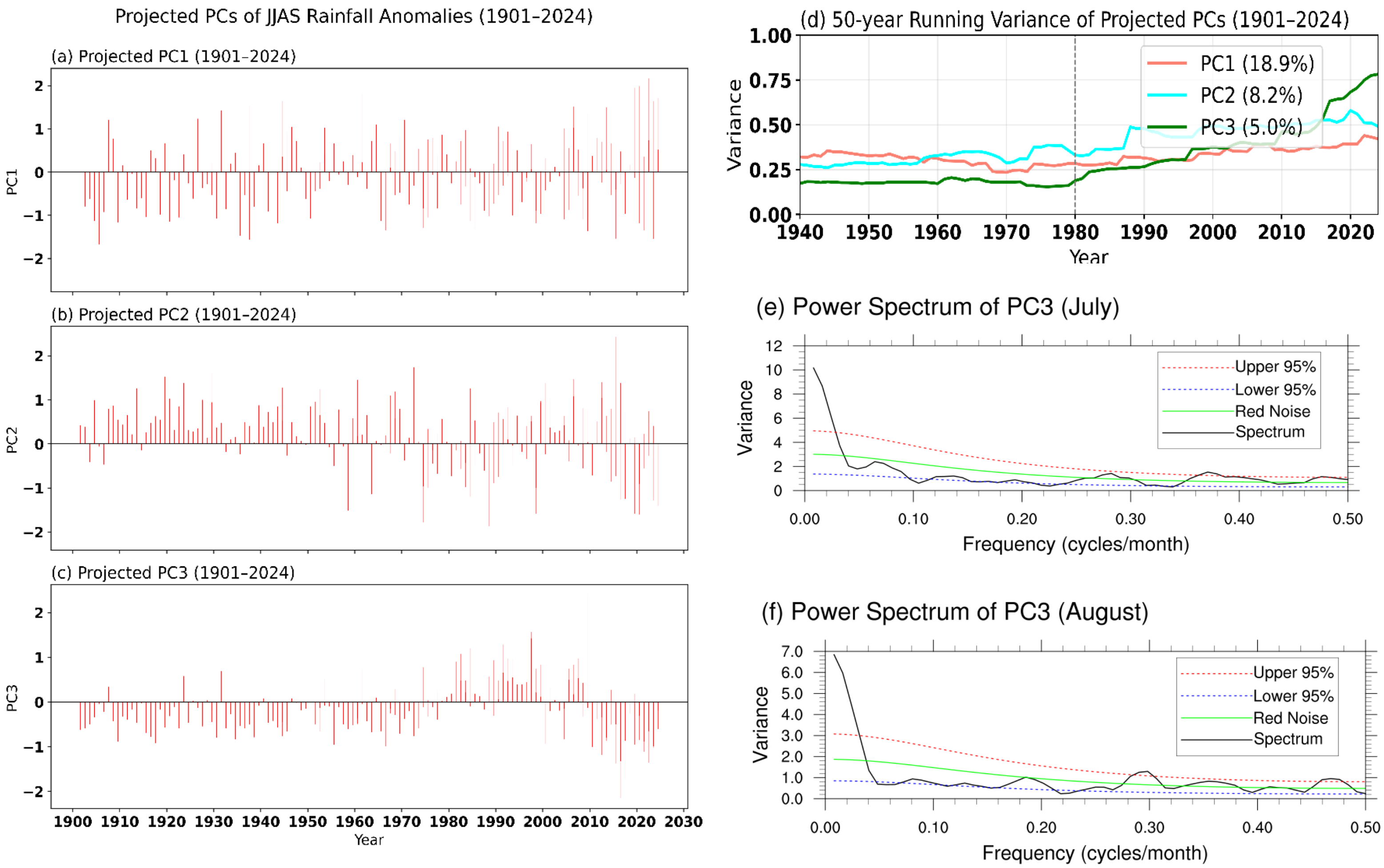

**Fig.4** The left panel shows PCs projected on to the computed PCs (shown in Fig.2) for 125 years (1901-2024) for July-August months (d) The 50-year running variance of the projected PC of the 3 dominant modes for 125 years (e) the power spectrum of projected PC 3 for July-August.

On analyzing the 3rd mode with a variance of 5.01% of rainfall, we can see a clear east-west dipole pattern (**Fig.2 (e)**). It shows a similar spatial pattern over the northwest, mainly across Gujarat and Rajasthan, and extends to southeast India through the peninsula, while the Gangetic plains, northeast regions, and southwest coast have opposite phase. This pattern contrasts with the usual monsoon rainfall distribution and aligns with the current trend of a westward shift in the monsoon rainfall. **Fig. 2**. (b), (d), and (f) represent the PC time series of the three EOF modes for rainfall anomalies, calculated on a monthly scale for July and August of each year. PC1 (**Fig.2b**) showed an interannual variability and did exhibit many long-term changes. Similarly, PC2 (**Fig.2d**) showed year-to-year variability and fewer long-term changes. PC3 (**Fig.2f**) shows an increased frequency of the positive phase in recent decades, reflecting a tendency for reduced rainfall in the east and enhanced rainfall in the west. When projected over the past century (**Fig. 4a-c), PC3 exhibits slow multi-decadal variability with a recent increase in its amplitude,** especially after 1980. This pattern is mainly linked to the east-west rainfall contrast in the monsoon zone. In contrast, PC1 and PC2 showed interannual variability, largely capturing the rainfall variations over the monsoon core region. Analysis of the 50-year running variance of the three PCs over time(**Fig.4d**) showed that PC3 exhibited a steep increase after 2000, whereas PC2 increased (insignificantly) and PC1 was stable. This indicates that the impact of Mode 3 on the monsoon distribution is increasing and there is an overall shift in the rainfall pattern. Spectrum analysis of this PC also reveals that modes 1 and 2 represent interannual or seasonal variability, whereas mode 3 has long-term (multi-decadal) variability (**Fig.4e** and **f**).

One of the major components of Indian monsoon rainfall is the vertical wind shear produced by two prominent jet streams over the Indian subcontinent. The tropical easterly jet at 200hPa generated from the large-scale anti-cyclonic circulation near the Tibetan Plateau and the low-level jet at 850hPa generated from the southwest Indian Ocean that brings large amounts of moisture to the Indian subcontinent (Findlater 1969) together form this wind shear. This can also be considered an inflow-outflow pattern that plays an important role in creating favorable conditions for monsoon rainfall.

The Webster Yang Index is a well-established measure through which we can quantify the vertical wind shear over the Indian region during the JJAS season. It is defined as the difference between zonal winds at 850 hPa and 200 hPa between 0-20˚N latitude and 40-110˚E longitude, which explains the monsoon variability provides a dynamic link between large-scale circulation patterns and monsoon rainfall, and also captures the changes in the monsoon circulation due to the effect of ENSO (Webster and Yang, 1992). A regression analysis of monsoon rainfall with this WYI (**Fig.5**) was performed for the July-August season, which gives an east-west dipole pattern, that is, the index has a positive influence on rainfall in the east over the Gangetic plains and northeast and also across the Western Ghats and southwest coastal regions, whereas it negatively affects the northwestern and south-eastern rainfall. Since this WYI has experienced an obvious decreasing tendency since 1972 and is associated with reduced summer precipitation around the Bay of Bengal (Zuo et al., 2013), this can be connected to the reduced rainfall in the eastern side and

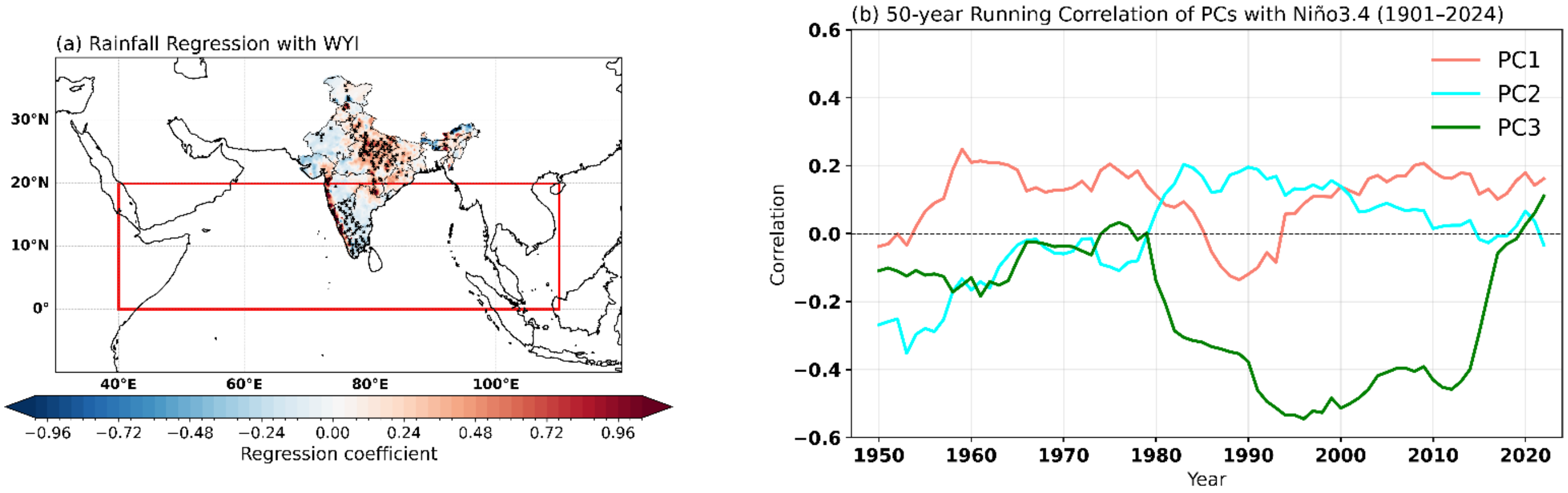

**Fig.5** (a) Regression map of Rainfall anomaly to reference index of Webster-Yang Index timeseries. It is the area averaged zonal wind shear ($U_{850}$ –$U_{200}$) season over the region 0-20˚N latitude and 40 to 110˚E longitude during July-August. Black dots represents areas with significant regression anomaly pattern. (b) 50-year running correlation of the 3 dominant PCs with SST (obtained from COBE_SST) anomaly over Nino3.4(5˚S - 5˚S, 170˚ - 120˚W) hence the recent shifts in the rainfall patterns from east to west. While weak circulation or reduced wind shear is associated with reduced rainfall availability over India during the monsoon, the spatial regression pattern shows that it is not uniform across the country. Some regions, particularly the monsoon core zones, have a positive effect, whereas certain regions, such as the northwest and southeastern peninsular regions, have a negative link to this wind shear. This exhibits spatial inhomogeneity in the rainfall distribution during the monsoon.

Furthermore, the WYI is significantly correlated with ENSO events; during El Niño, the monsoon weakens, and during La Niña, it strengthens. This suggests that the mode 3 pattern might be influenced by ENSO events. To further understand this relationship a 50-year running correlation between the projected PCs of the 3 dominant modes and SST anomaly over nino3.4 (5˚S - 5˚S, 170˚ -120˚W) is computed (**Fig.5b**). PC1 and PC2 were not strongly associated with the index, whereas PC3 exhibited an interesting relationship. After the 1980s, ENSO and mode 3 were significantly correlated (around -0.43) and during the recent decade, this correlation weakened. This suggests that the recent shift in the monsoon rainfall pattern might be a consequence of the ENSO-Monsoon decoupling.

The similarity of the regression map (**Fg.5**) and the mode 3 EOF pattern (Fig.2c) shows that the pattern is not only a statistical construct of rainfall variability, but also represents a dynamically meaningful feature. Furthermore, the principal component of mode 3 has a 0.55 correlation with the index, which is significant at a 95% confidence level based on a Student's t-test. Therefore, both analyses together show the existence of an east-west dipole pattern in monsoon rainfall distribution which is different from the classical homogenous monsoon pattern.

## East–West Rainfall Asymmetry: Observed pattern and features

The pronounced shift in the rainfall distribution pattern over India can be characterized by analyzing the difference in the standardized rainfall anomaly ($\mathbf{A_{we}}$) between two distinct regions of the country, western and eastern India, $\mathbf{A_{we}} = (\mathbf{A_w} - \mathbf{A_e})$, as defined in Section **2.2**. Here, the rainfall gradient is calculated as a west-east difference, and it revealed a significant positive trend in the last decades, particularly after 2010, indicating an increase in rainfall over the western region relative to the east (**Fig. 6a**). **Fig.7** shows the composite rainfall anomaly patterns for the strong negative and positive phases of this rainfall gradient, constructed from years below the 20th percentile (approximately) and above the 80th percentile (approximately. The composites show that during positive years, the western region receives more rainfall, whereas the eastern region experiences lower rainfall, and the pattern reverses in negative years. This clear east–west contrast supports the emergence of a dipole pattern in the rainfall distribution during the monsoon.

.

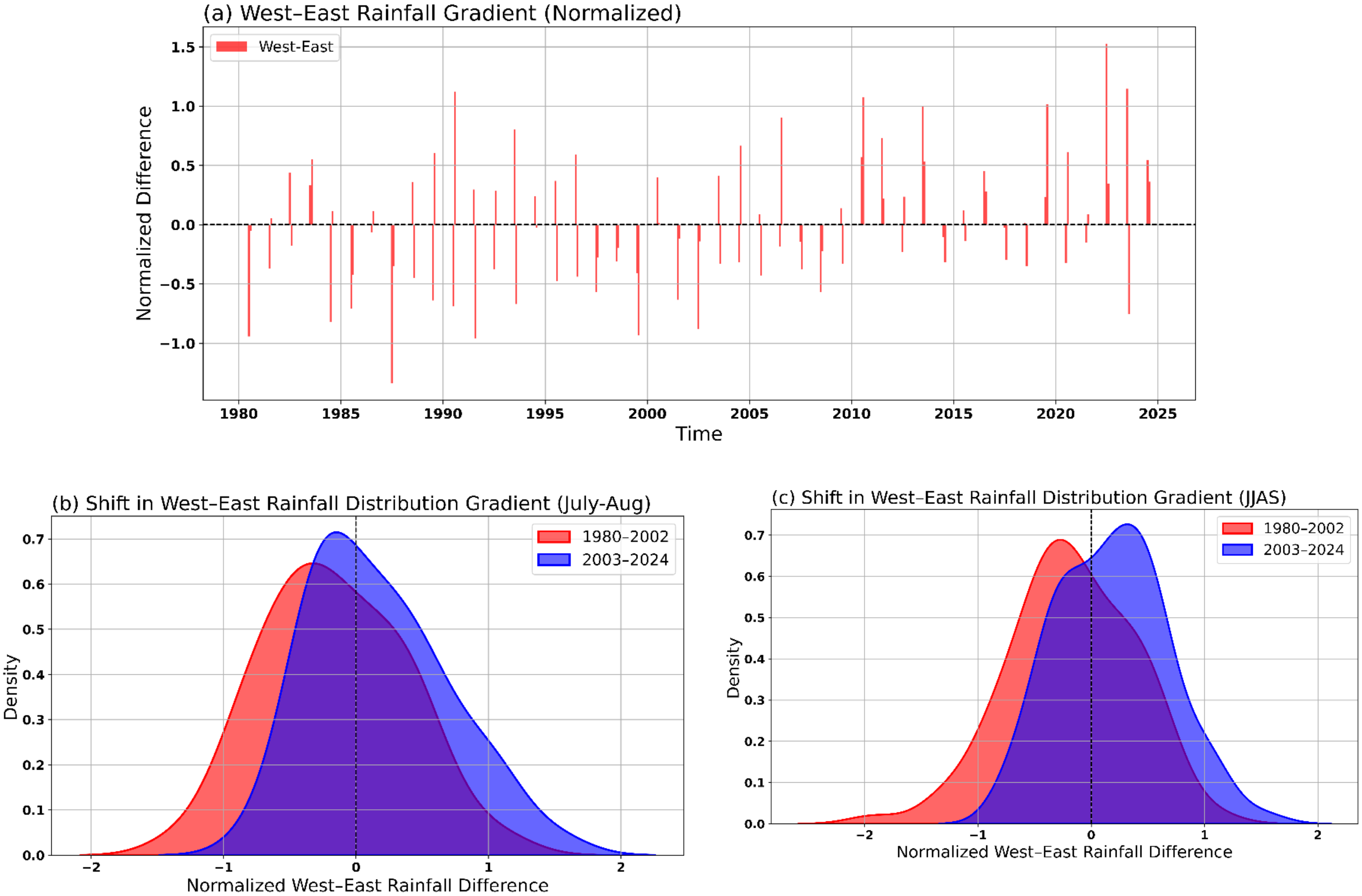

**Fig.6** Monthly variation of the normalized rainfall anomaly gradient (**$A_w$ - $A_e$**) Index for July–August (b) the PDF representation of the anomaly difference for July-August for 2 periods (red represents the earlier decades 1980 to 2002 and blue represent the recent decade 2003 to 2024) (c) PDF representation of the anomaly difference for JJAS season for 2 periods (red represents the earlier decades 1980 to 2002 and blue represent the recent decade 2003 to 2024).

A Probability Density Function (PDF) analysis of this standardized rainfall anomaly gradient (**$A_{we}$**) for the two periods from 1980 to 2002 and 2003 to 2024 is shown in **Fig.6** b and c. This reveals a significant positive shift in rainfall during recent years compared to earlier periods with a 95% confidence level according to the Student's t-test. In the earlier period, the PDF distribution was skewed to negative values, indicating that more rainfall was received in the east. In contrast, during the recent period, the distribution shifted to the right, reflecting a westward shift in rainfall availability. This shift can be observed for the July-August months and for the entire JJAS season

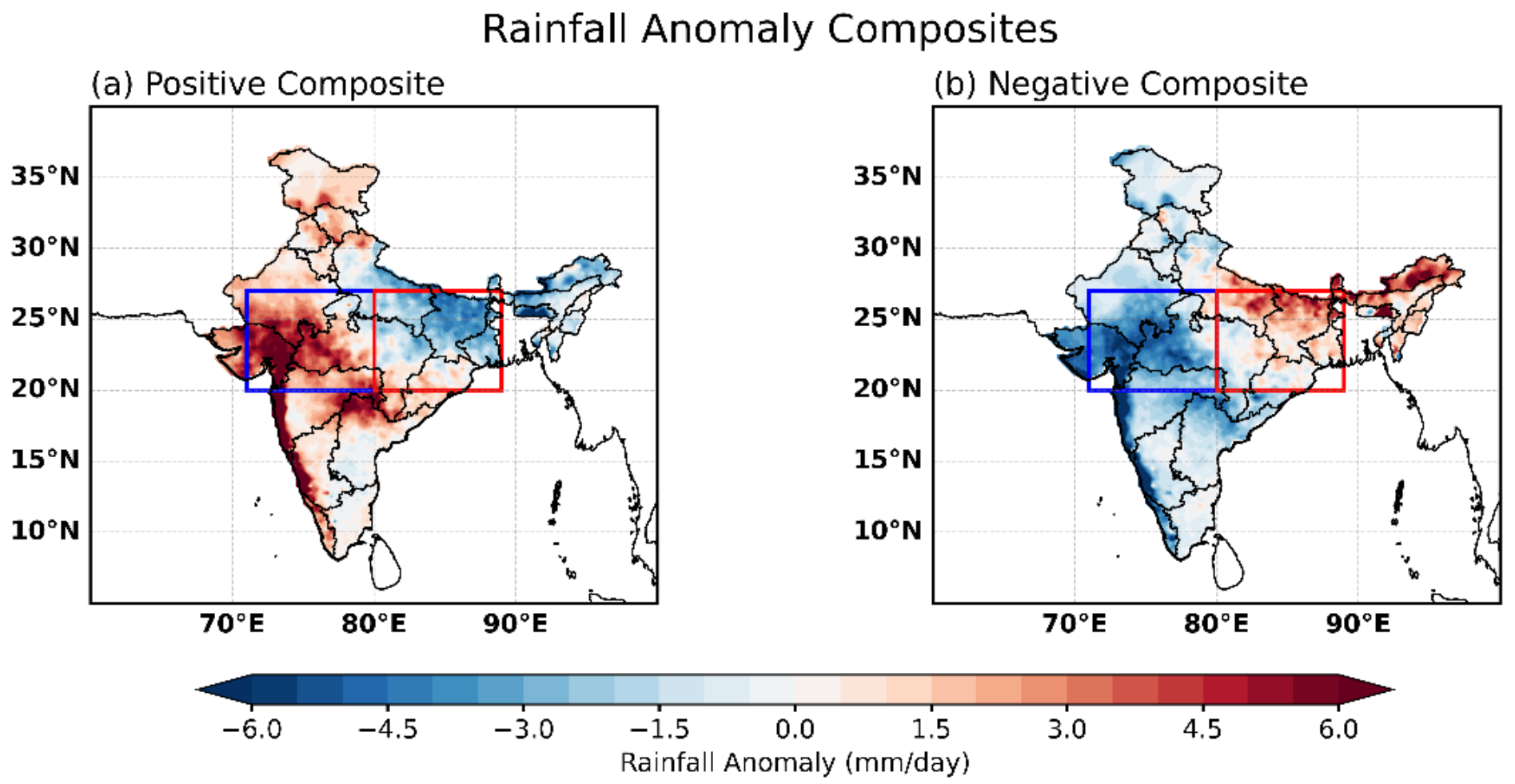

**Fig.7** Composite rainfall anomalies for July-August(mm/day) over 1980–2024 based on the rainfall gradient (**$A_{we}$=$A_w$ - $A_e$**). The left panel shows the lowest 10 years (~bottom 20th percentile), and the right panel shows the highest 10 years (~top 80th percentile) selected by arranging the index value in descending order. Boxes indicate the west (71°–80°E, 20°–27°N) and east (80°–89°E, 20°–27°N) regions used to compute the index.

(**Fig.6(b & c)**). The shift during the JJAS season reveals a more pronounced displacement of the distribution, which indicates how this shift is happening prominently for the full monsoon season. These observed features highlight the spatial asymmetry in the monsoon rainfall distribution and its recent westward shift.

## 3.2 East–West Rainfall Asymmetry: Dynamical Drivers

Even though there exist a dipole mode of the monsoon rainfall pattern represented by EOF mode 3, it is not the physically dominant mode of variability as per the EOF analysis. However, in recent years, the rainfall anomalies have shown an increased resemblance to this mode (**Fig.1**), indicating the emergence of an east-west dipole pattern in the rainfall distribution. This emerging shift in rainfall distribution reflects the underlying large-scale changes in monsoon dynamics. Therefore, a comprehensive analysis is required to understand the drivers of this transition. It is essential to understand how large-scale circulation, basin-scale pressure, and moisture influence the emerging dipole rainfall distribution pattern. In the next few paragraphs, we describe some important dynamic features.

*(a) SLP*

In this study, the contributing factors influencing rainfall variability in the eastern and western regions of India were analyzed independently. **Fig.8a** shows the spatial trend of SLP for the period 1980 - 2024, and it clearly shows how the pressure decreases significantly in the equatorial Indian Ocean region and over the Arabian Sea. To better understand the temporal evolution, trend analysis was further performed for two subperiods: 1980 to 2002 and 2003 to 2024 (**Fig.8 b, c**). In earlier times, there was an increasing trend in the Arabian Sea and a decreasing SLP over the Bay of Bengal; during the later period, this trend was reversed. Furthermore, the PDF distribution of the SLP over the Arabian Sea between 5˚-20˚ N Latitude and 57˚-72˚E longitude also shows a leftward shift over recent decades, consistent with the overall decreasing trend, whereas it was skewed towards positive values in the earlier period (**Fig.9a**). This observed reduction in SLP over the northern Arabian Sea enhanced the low-level convergence. This is likely to facilitate the accumulation of moisture owing to advective transport, resulting in subsequent cloud development.

The Bay of Bengal region, which that is directly associated with rainfall in northeastern India, was also studied (**Fig.8a**). It reveals an insignificant increase in SLP along the head of the Bay of Bengal, and over the southern region, there is a significantly decreasing trend. **Fig.9(c)** presents the PDF distribution of the Bay of Bengal region between 5˚-20˚ N latitude and 80-95˚ E SLP, which indicates no significant shift in recent years as compared to the earlier. Whereas a marked decrease in SLP is observed over the Arabian Sea, which creates favorable conditions for rainfall.

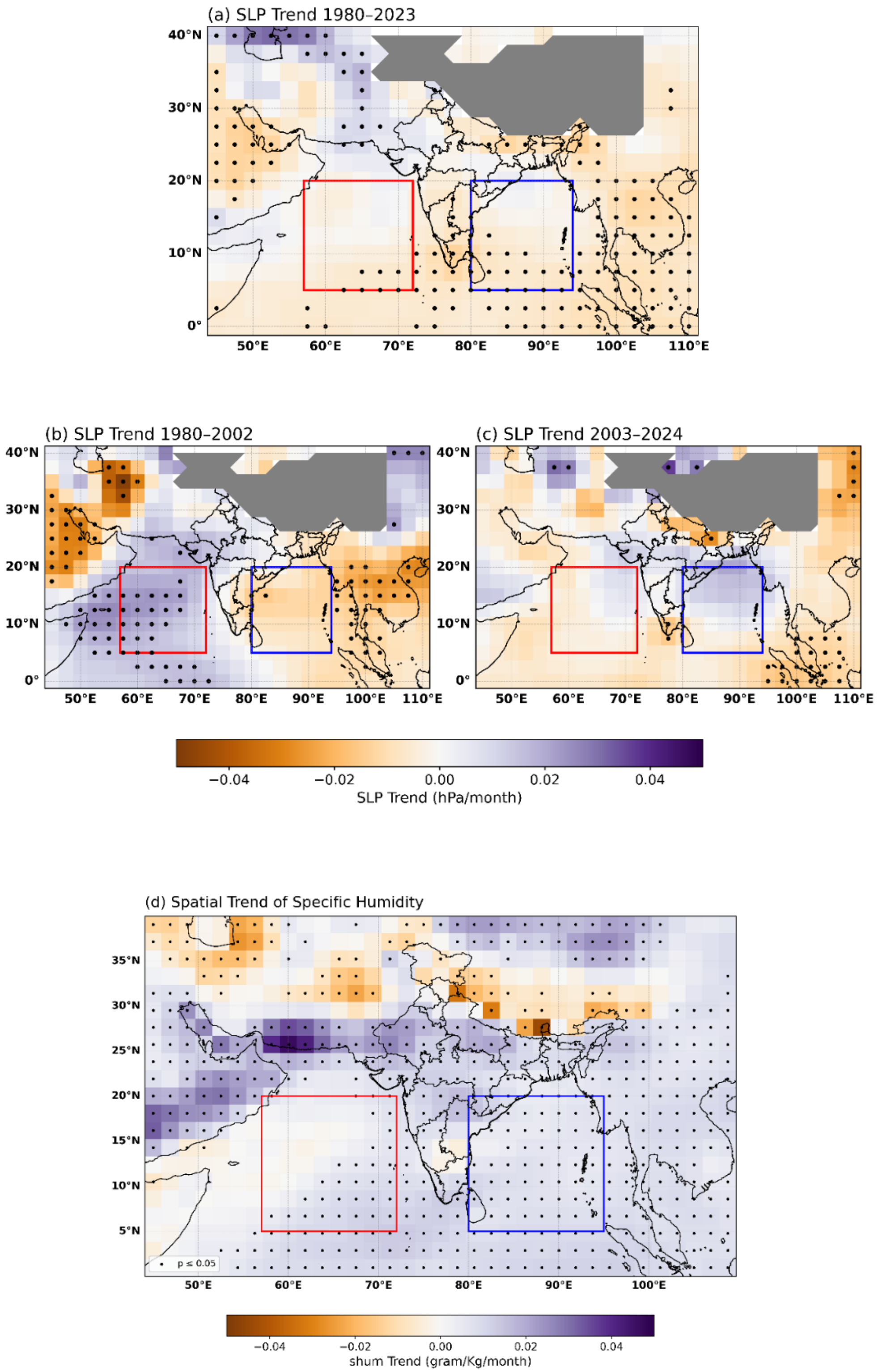

**Fig.8** (a) July-August monthly trend analysis for SLP for the time period 1980 to 2024 (b) July-August monthly trend analysis for SLP for the time period 1980 to 2002 (c) July-August seasonal trend analysis for SLP for the time period 2003 to 2024 (High mountain regions are masked) (d)July-August seasonal trend analysis for specific humidity for the time period 1980 to 2024, the black dots represent areas of 95% confidence level.

### (*b) Specific Humidity*

To understand the moisture transport to northwest and northeast India, the changes in specific humidity were also analyzed (**Fig.8(d)**). Over the Arabian Sea, it exhibited a pronounced increasing trend, reflecting an elevated atmospheric moisture content. This was likely driven by intensified evaporation due to ocean warming. The PDF distribution of the specific humidity (**Fig.9b**) further reveals a positive shift in the humidity distribution over the Arabian Sea, reflecting the enhanced moisture availability in this region. Even though the humidity increases over both the Arabian Sea and the Bay of Bengal (**Fig.9**c and d), the rainfall shifted westward. This might be due to the concurrent rise in SLP, which inhibits efficient cloud development over the region and maintains dynamic stability. This may have hindered enhanced precipitation over the associated northeastern regions.

In contrast, over the Arabian Sea, the SLP showed a decreasing trend and the moisture content increased. Together, these conditions create a dynamic environment favorable for the genesis and intensification of monsoon depressions and cyclonic systems over the Arabian Sea. Recent studies have demonstrated that the frequency of monsoon depressions over the northern Arabian Sea has nearly doubled. During 2001–2022 compared to 1981–2000, a trend attributed to changes in both dynamic and thermodynamic conditions, including elevated sea surface temperatures and enhanced moisture flux convergence (Chilukoti et al., 2024). These factors, together with favorable circulation patterns, promote enhanced convective activity and precipitation over northwestern India. While depressions contribute to enhanced rainfall over western India, they also act as evidence of broader dynamic changes, such as the lowered SLP, increased atmospheric moisture, and strengthened circulation. Together these factors together act as the fundamental drivers behind the observed westward shift in monsoon rainfall.

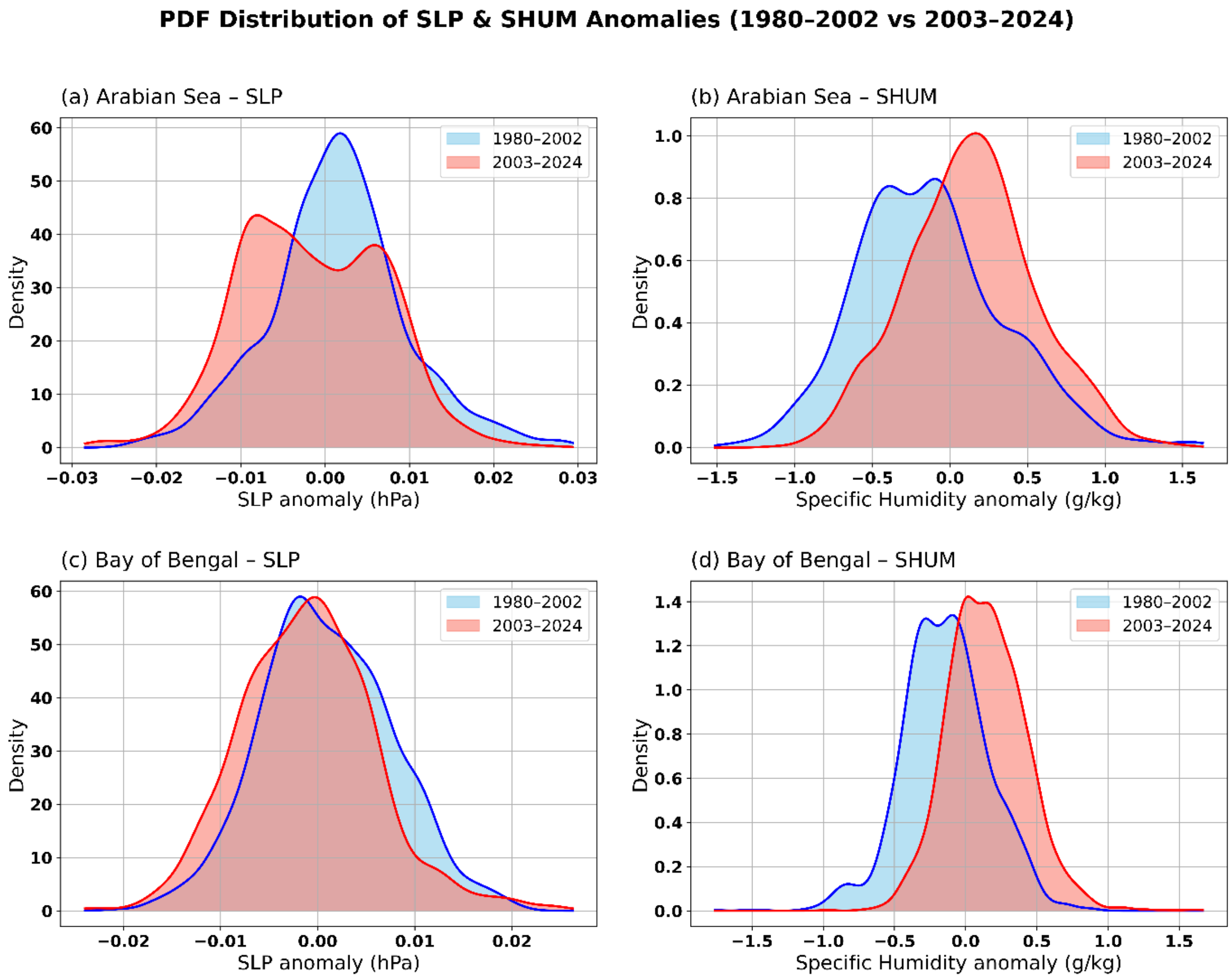

**Fig.9** PDF representation of (a) SLP (hPa) over Arabian Sea (b) specific humidity (grams/Kg) over Arabian Sea between 5˚-20˚ N Latitude and 57˚-72˚E longitude (c) SLP (hPa) over Bay of Bengal (d) specific humidity (grams/Kg) over Bay of Bengal between 5˚-20˚ N latitude and 80-95˚E longitude for 2 periods (red represents the earlier decades 1980 to 2002 and blue represent the recent decade 2003 to 2024). The analysis is done on monthly scale for July-August.

## *(c) Wind*

It is well known that the wind pattern plays a central role in determining rainfall distribution, as moisture transport by atmospheric circulation dictates which regions receive the moisture necessary for precipitation. The current analysis shows a spatial trend of zonal wind at 850hPa, showing an increase near the equator along the Somali coast but a decreasing trend over the southern Indian mainland which is directed towards the eastern region(**Fig.10a**). Particularly, over the Arabian Sea: at the southwestern coast where the jet typically enters via Kerala and Karnataka,

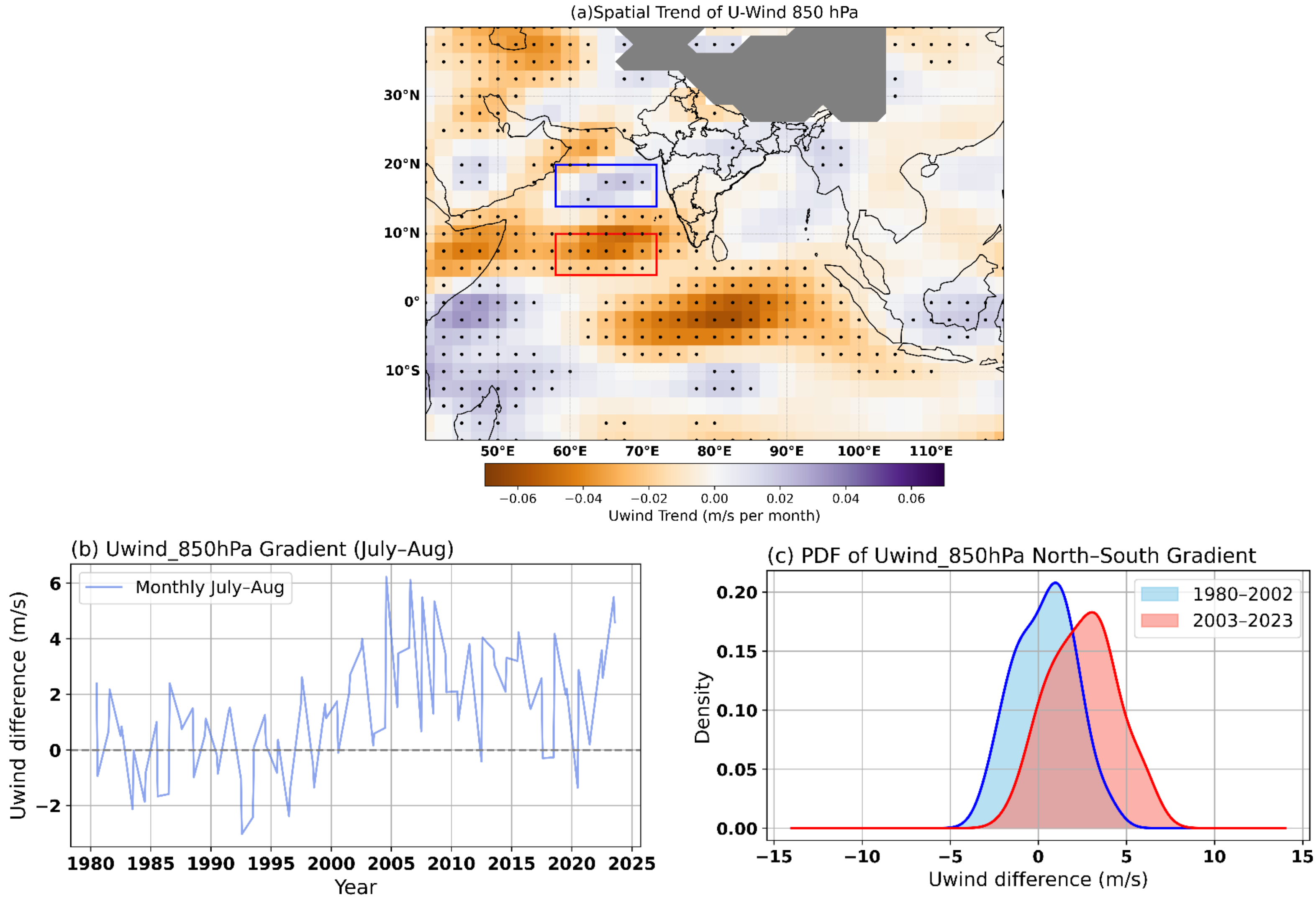

**Fig.10** (a) July-August monthly spatial trend analysis for zonal wind(u)(m/s) for the time period 1980 to 2024, the black dots represent areas of 95% confidence level. The rectangle box shows the region considered for calculating the North-South U-wind gradient in the Arabian Sea. (High mountain regions are masked) (b) Time series of U-wind gradient (north-south) for July-August months (c) The PDF distribution of the uwind gradient for two timeperiod:1980-2002 and 2003-2024.

triggering monsoon onset, a significant decrease in zonal wind(u) was observed. In contrast, the in northern Arabian Sea near the northwest coast of India (Gujarat, Maharashtra) the zonal wind exhibited a substantial increasing trend. Monthly variability in the North-South gradient of zonal wind is calculated by considering two boxes; North box between 14-20˚N Latitude and 60-71˚E longitude and south box between a 0-10˚N latitude and 58-70˚E longitude, this confirms intensification of the low-level jet over northern Arabian Sea, and PDF analysis of this gradient further evidences a poleward shift of the jet in the recent decades (**Fig.10c**). The LLJ diverges into two branches in the Arabian Sea, the Northern branch intersects the west coast of India and the southern branch of the split moves eastward just south of India (Krishnamurti et al., 1976). This configuration favors increased rainfall over northwestern India by directing more moisture towards this region, while reducing moisture transport and precipitation on the eastern side. Thus, the PDF analysis indicates the recent intensification of the northern branch of the jet in the Arabian Sea and a reduction in the southern branch that moves further east and brings more moisture to this region.

The change in this wind pattern can also be considered long-term evidence of Indian Ocean warming and changes in the associated components. Joseph et al., 2024 demonstrated that the strengthening of the cross-equatorial monsoon winds is due to the rapid warming of the Indian Ocean and enhanced Pacific Ocean trade winds, which result from the poleward shift and expansion of the Hadley cell. These strengthened winds boosted the latent heat flux (evaporation), leading to increased moisture transport to Northwest India. This zonal wind pattern shift at the lower level can also explain the decrease in rainfall in the eastern region despite the generation of more moisture in the Bay of Bengal due to the ocean warming.

### *(d) SST*

Climatologically, the Bay of Bengal has a higher sea surface temperature (SST) amplitude than the Arabian Sea (Prasanna Kumar and Narvekar, 2005). Studies have shown that the Indian Ocean exhibits a robust basinwide sea surface temperature (SST) warming during the twentieth century, which has affected the hydrological cycle, atmospheric circulation, and global climate change (Dong et al., 2014). However, in recent years, studies have shown a significant slowdown in SST warming in the Bay of Bengal compared to the Arabian Sea (Albert et al., 2023). **Fig.11** shows the SST warming difference over the Arabian Sea and Bay of Bengal between 1980-2002 and 2003-

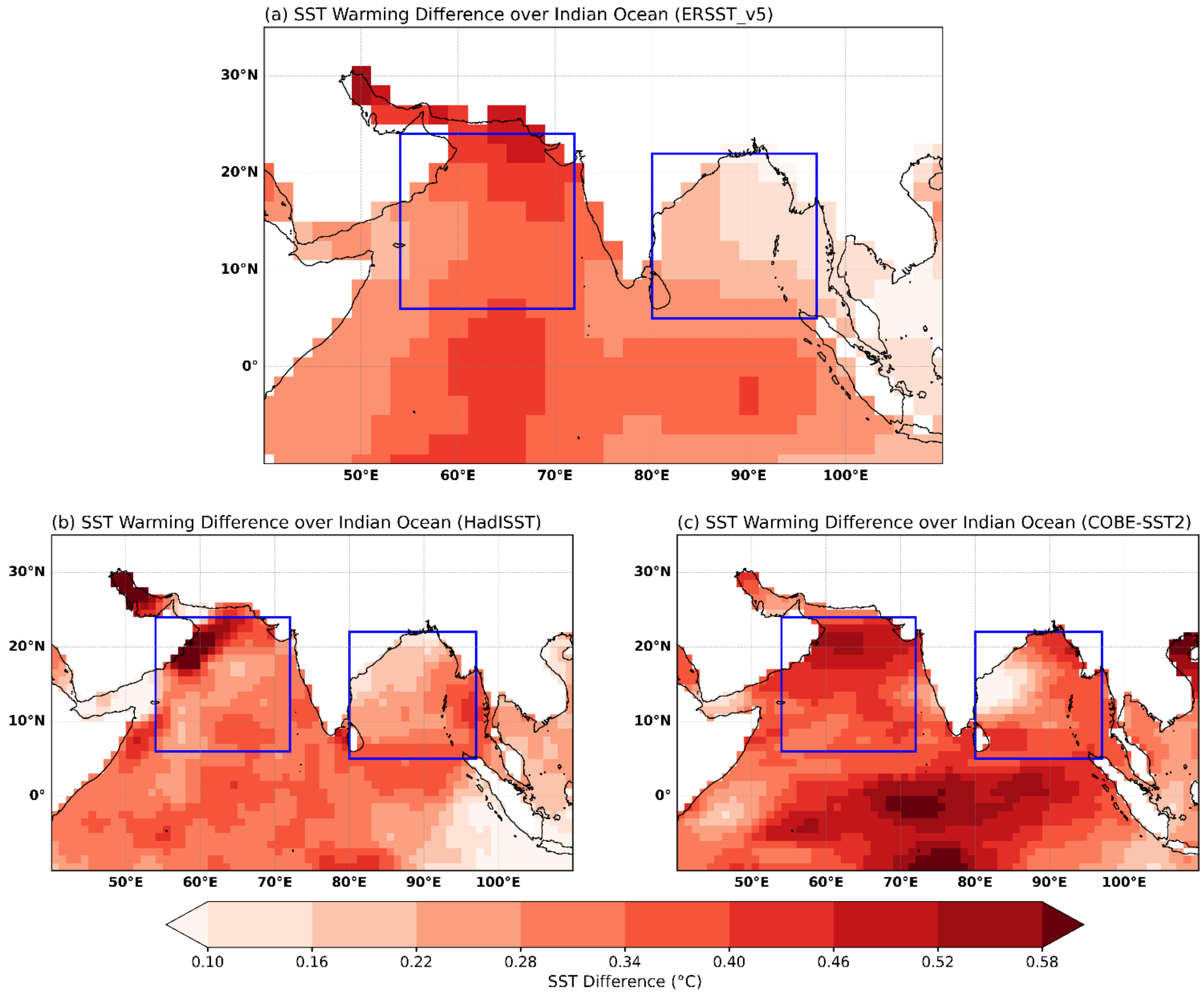

**Fig. 11** Spatial plot of SST warming difference (degree Celsius) between two periods recent period (2024-2003) and earlier period (1980-2002) from (a)ER_SST_v5 dataset (b) HadISST data (c) COBE_SST2 data for July-August. The rectangle box shows the region considered for Arabian Sea and Bay of Bengal for the comparison study.

2024 based on three different SST datasets: ERSST_V5 (**Fig.11a**), HadISST (**Fig.11a**), and COBE_SST (**Fig.11a**). This indicates that warming is more pronounced over the Arabian Sea than over the Bay of Bengal. Although COBE_SST slightly overestimates warming in the equatorial Indian Ocean compared to other datasets, this region falls outside the scope of our study. Furthermore, the PDF distribution of this SST warming difference shows that the Arabian Sea (6˚-24˚ N, 54˚ -72˚ E) has a greater warming over the past decade, with a warming of around 0.4˚C, whereas over the Bay of Bengal (5˚-23˚ N, 80˚-96˚ E), it is approximately 0.2˚C. Three different SST datasets, including ERSST_V5, HadISST, and COBE_SST, verify that warming over the Arabian Sea is more than that over the Bay of Bengal (**Fig.12**) in recent years. This clearly shows that the Arabian Sea is warmer than previous decade, which influences SLP, and moisture availability is the key factor for rainfall over the northwest. This rapid change in SST over the Arabian Sea and other dynamic factors together cause profound changes in the spatial rainfall distribution over India.

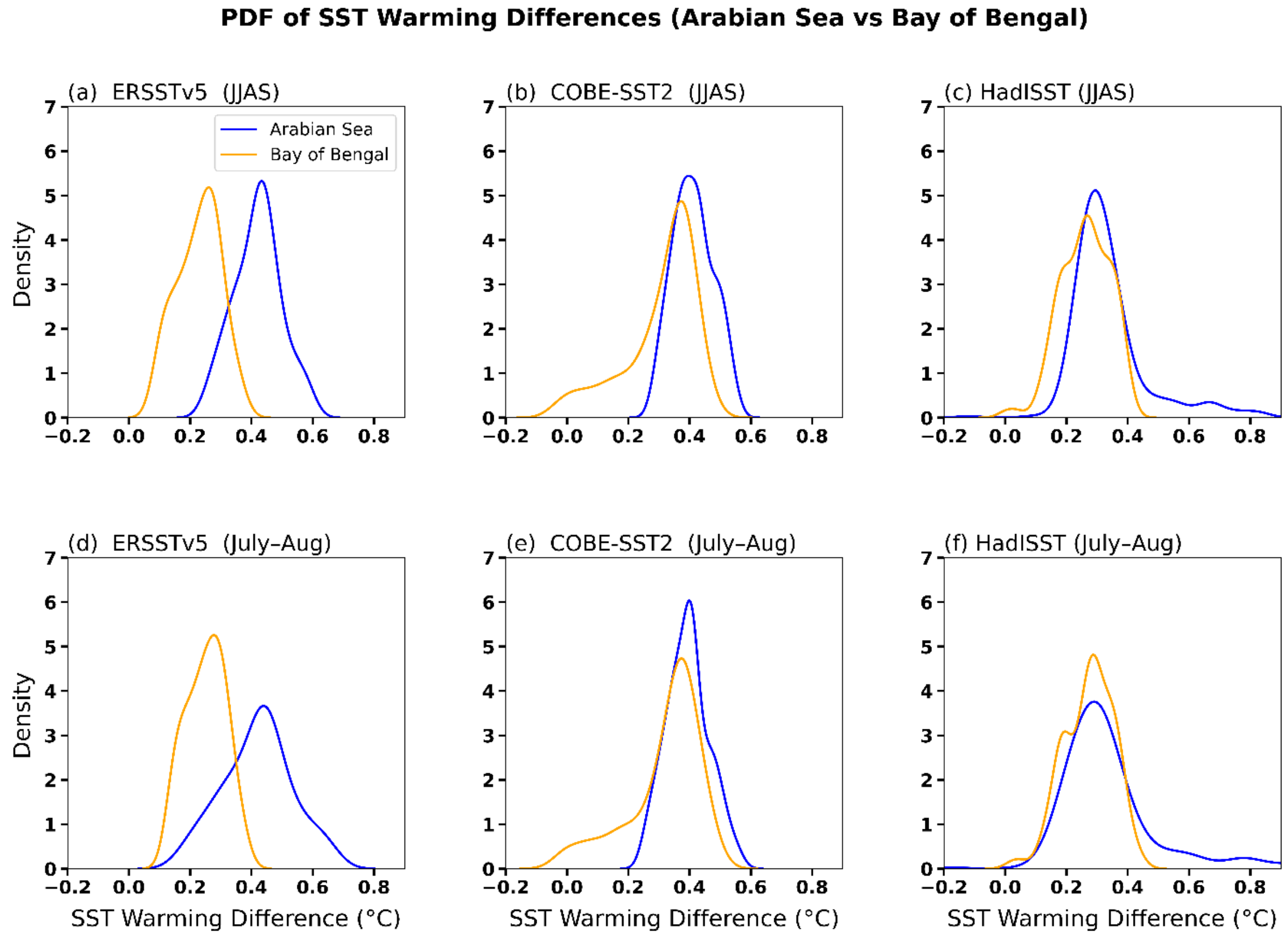

**Fig 12**. PDF distribution of SST warming difference between two periods recent (2003 to 2024) and earlier (1980 to 2002) for JJAS season and July-August months. The orange line represents Bay of Bengal region between 5˚-23 ˚ N latitude and 80 ˚-96 ˚ E longitude and blue line represent Arabian Sea over the box between 6 ˚- 24 ˚ N latitude and 54˚ -72 ˚ E longitude.

# 4. Discussions

This study investigates the probable dynamical association of the changing spatial pattern of Indian Summer Monsoon monthly rainfall and provides evidence of a clear east–west dipole emerging over India on a monthly to seasonal scale, particularly during the core monsoon months of July and August. However, this rainfall pattern was, not a new. The EOF based on long-term monthly data shows that this mode exists as a pattern of rainfall variability over India, which has a lower variance. This was the third rainfall mode. The EOF is physically reproduced by the spatial pattern of the regression of rainfall anomalies with respect to the Webster–Yang Index(**Fig.5**). The regression pattern in the figure further supports the existence of a dynamic linkage in this dipole pattern. The trend in the variance analysis in **Fig.4** (b),(d), and (f) shows that earlier variability with respect to this mode was masked by the other dominant modes. However in recent decades, , the increasing amplitude of PC3 and the corresponding rainfall anomalies indicate that this dipole pattern is becoming more prominent (increased variance compared to the others).

The significant relationship between Mode 3 PC and WYI suggests a potential link between Mode 3 and ENSO. A 50-year running correlation between ENSO and the principal components revealed that PC3 is associated with ENSO (**Fig.5b**). During earlier times (before the 1970s), it had a weak association with ENSO but after the 1980s, it became significant (around -0.43). Furthermore, after the 2000s, the relationship weakened and became statistically insignificant, whereas this mode continued to intensify. This suggests that the earlier variability in PC3 was influenced by ENSO, but in recent decades, it has become decoupled from ENSO. However, this increased variance now appears to be governed by regional factors. For example, Arabian Sea warming and shifts in the low-level jet now influence the rainfall over the western part of the country. This observation highlights that the recent strengthening of the east–west dipole pattern is the result of the weakening of the ENSO-Monsoon relationship, along with the changing regional dynamics as a consequence of global warming.

The study also examines the associated regional dynamical shifts, particularly over the Arabian Sea and Bay of Bengal. Probability density function (PDF) analysis shows that in recent decades, there has been a decrease in sea level pressure (SLP) over the Arabian Sea, while conditions over the Bay of Bengal have remained largely unchanged (Fig. 9a and **b**). This decline in SLP, along with an increase in moisture content, enhanced convection and cloud formation over the Arabian Sea, leading to increased rainfall over northwestern India. The increasing frequency of depressions over the Arabian Sea in recent years can be considered a consequence of this shift.

The circulation patterns also reflect these changes. The low-level jet at 850 hPa, which is one of the crucial factors of the Indian monsoon, shifted northward. The southern branch of the jet shows a decreasing trend, whereas the northern branch was strengthened (**Fig.10a)**. This shift in the key moisture transport mechanism of the monsoon directly alters the rainfall distribution across the subcontinent. Further investigation of the SST variation of the Arabian Sea and Bay of Bengal shows that with respect to the earlier decades over the recent period, the Arabian Sea has experienced more warming than the Bay of Bengal (**Fig.12**). This change in SST warming contributes to a dip in SLP over the Arabian Sea, increased moisture flux, and intensified precipitation over northwestern India. The shift in the low-level jet (Fig.10c) also contributed to increased rainfall over the northwestern parts of the country.

The analysis reflects two things: (a) there is a shift in the variance trend in the mode-3 rainfall pattern, which shows an increasing trend in variance; and (b) it is consistent with local dynamical features in the Indian Ocean (BoB and Arabian Sea) and its likely association with the weakening of the ENSO-monsoon relationship. Although the analysis did not delve into low-frequency decadal changes in the monsoon, the current analysis showed a multi-decadal change associated with a shift in the spatial monsoon pattern. This shift in rainfall pattern has far more implications. This increase in east-west asymmetry in the rainfall distribution affects regional water resources, increasing flood risk in the northwest while intensifying drought conditions in the east. Such variability poses critical challenges for water management, agriculture, and food security, all of which are highly dependent on monsoon rainfall. Therefore, the current analysis will help in understanding their shifting patterns and will provide crucial information for developing climate-resilient agricultural strategies and sustainable resource management in the coming decades.

## Acknowledgments

*RC* and *AS* acknowledge the Director, Indian Institute of Tropical Meteorology (IITM), for academic support. *AS* was reported as part of this work as an MSc internship. AS acknowledges the support from Prof. K Ashok, Centre for Earth, Ocean and Atmospheric Sciences (CEOAS), School of Physics, University of Hyderabad, India, for valuable discussions during the MSc internship period.

## Author Contributions

RC conceptualized and designed the study and assisted in drafting the manuscript. AS did the scientific computations, plotting, and drafting of the initial version of the manuscript.

## Data Availability

The study uses free data sets from their respective websites, rainfall data from Indian Meteorological Department (https://imdpune.gov.in/cmpg/Griddata/Rainfall_25_NetCDF.html), zonal(u) and meridional(v) wind, sea level pressure(SLP) and specific humidity is obtained from National Centers for Environmental Prediction NCEP (https://psl.noaa.gov/data/gridded/data.ncep.reanalysis.html), SST data was obtained from COBE_SST 2 and Sea Ice (https://psl.noaa.gov/data/gridded/data.cobe2.html) , NOAA Extended Reconstructed SST V5 (https://psl.noaa.gov/data/gridded/data.noaa.ersst.v5.html) and from Hadley Centre Sea Ice and Sea Surface Temperature (https://www.metoffice.gov.uk/hadobs/hadisst/data/download.html).

## Competing Interests

The authors declare there is no competing interest

## Funding

There is no funding available to be reported

## Ethical Compliance

Not applicable